\documentclass[prd,preprint,superscriptaddress,floatfix,eqsecnum,nofootinbib,12pt]{revtex4}
\usepackage{hyperref}
\usepackage{amssymb,amsmath,amsthm,graphicx,ulem}
\pdfoutput=1
\usepackage{graphicx,subfig}
\usepackage{ulem}
\usepackage{epsfig}
\usepackage{amsmath}
\usepackage{amsfonts}
\usepackage{amssymb}
\usepackage[usenames]{color}
\usepackage[letterpaper,left=2.cm,right=2.cm,top=2.5cm,bottom=2.5cm]{geometry}
\usepackage[T1]{fontenc}
\usepackage{float}

\newcommand{\beq}{\begin{equation}}
\newcommand{\eeq}{\end{equation}}
\newcommand{\be}{\begin{equation}}
\newcommand{\ee}{\end{equation}}
\newcommand{\beqa}{\begin{eqnarray}}
\newcommand{\eeqa}{\end{eqnarray}}
\newcommand{\beqar}{\begin{eqnarray*}}
\newcommand{\eeqar}{\end{eqnarray*}}
\newcommand{\bea}{\begin{eqnarray}}
\newcommand{\eea}{\end{eqnarray}}

\newcommand{\dd}{\textrm{d}}

\newcommand{\nn}\nonumber

\def\di{\mathrm{d}}

\newcommand{\beginsupplement}{%
        \setcounter{table}{0}
        \renewcommand{\thetable}{S\arabic{table}}%
        \setcounter{figure}{0}
        \renewcommand{\thefigure}{S\arabic{figure}}%
     }

\begin{document}

\title{Damped perturbations in the no-boundary state}

\author{J. Diaz Dorronsoro}
\affiliation{Institute for Theoretical Physics, KU Leuven, 3001 Leuven, Belgium}
\author{J. J. Halliwell}
\affiliation{Blackett Laboratory, Imperial College, London SW7 2BZ, UK}
\author{J. B. Hartle}
\affiliation{Department of Physics, UCSB, Santa Barbara, CA 93106, USA}
\author{T. Hertog}
\affiliation{Institute for Theoretical Physics, KU Leuven, 3001 Leuven, Belgium}
\author{O. Janssen}
\affiliation{Center for Cosmology and Particle Physics, NYU, NY 10003, USA}
\author{Y. Vreys}
\affiliation{Institute for Theoretical Physics, KU Leuven, 3001 Leuven, Belgium}

\begin{abstract}
\noindent
We evaluate the no-boundary path integral exactly in a Bianchi IX minisuperspace with two scale factors. In this model the no-boundary proposal can be implemented by requiring one scale factor to be zero initially together with a judiciously chosen regularity condition on the momentum conjugate to the second scale factor. Taking into account the non-linear backreaction of the perturbations we recover the predictions of the original semiclassical no-boundary proposal. In particular we find that large perturbations are strongly damped, consistent with vacuum state wave functions. \\

\begin{centerline}
{\it Dedicated to the memory of Stephen Hawking 1942 -- 2018}
\end{centerline}
\begin{centerline}
{\it Friend, teacher, colleague, and source of inspiration for many years to come}
\end{centerline}
\end{abstract}

\maketitle

\section{Introduction} \label{introsec}
\noindent A fundamental theory of our quantum universe consists of a theory of its dynamics and a theory of its quantum state -- a wave function of the universe. The no-boundary wave function (NBWF) of the universe \cite{HH1983,Hawking1983} is perhaps the most explored candidate for the theory of the state. In simple dynamical models it successfully predicts important features of our observed universe such as the existence of classical histories \cite{HH1983,Hawking1983,HHH2008}, an early period of inflation \cite{HHH2008,Hartle:2007gi}, a nearly-Gaussian spectrum of primordial density fluctuations \cite{Halliwell:1984eu,Hartle:2010vi,Hertog:2013mra}, a physical arrow of time \cite{Hawking:1993tu,HH2011}, etc.  

A wave function of a closed universe is a functional of the three metric $h_{ij}(x)$ and field configuration $\phi(x)$ on a spacelike three-surface $\Sigma$. It is fair to say that the above agreement of the NBWF with observation has been mostly obtained in minisuperspace models that only explore a limited nearly homogeneous and isotropic range of configurations, and in the semiclassical approximation only. In this approximation the no-boundary proposal for the state amounts to the selection of a particular set of saddle points of the action of gravity coupled to matter fields. Nevertheless the NBWF was originally motivated by a Euclidean functional integral construction.\footnote{The oft-used terms ``Euclidean'' and ``Lorentzian'' path integrals are only roughly indicative. The integration is generally over complex contours \cite{Halliwell:1990qr}.}

In this paper we verify the validity of the no-boundary idea by evaluating the no-boundary path integral exactly in a Bianchi IX-type minisuperspace model. The squashed three-spheres of this model are homogeneous but can have significant deviations from isotropy. Classically regularity of the no-boundary saddle points implies constraints on the metric and its first derivatives. These enter as variables and conjugate momenta in the quantum theory. In the anisotropic minisuperspace, which has two scale factors, we show that a proper implementation of the no-boundary idea as a functional integral is obtained by taking one scale factor to be zero initially together with a judiciously chosen regularity condition on the momentum conjugate to the second scale factor. These conditions imply that the classical configuration which dominates the path integral in the semiclassical limit is regular everywhere and has both scale factors equal to zero initially. The resulting normalizable quantum state predicts that both small and large deviations from isotropy are damped, in correspondence with previous considerations of the semiclassical Hartle-Hawking state in similar models \cite{HAWKING198483,WRIGHT1985115,PhysRevD.31.3073}. By contrast, alternative implementations of the no-boundary idea in this model fail to specify a well-defined state.

\newpage
\section{Biaxial Bianchi IX Minisuperspace} \label{Sec-pq}
\noindent We consider a homogeneous but anisotropic minisuperspace approximation to gravity coupled to a positive cosmological constant $2 \pi^2 \Lambda$ and no matter fields. The classical histories in this minisuperspace are known as biaxial Bianchi IX cosmologies which are non-linear versions of the lowest $n=2$ gravitational wave mode perturbation of de Sitter space. We write the metric of this minisuperspace model as
\begin{equation} \label{metricansatz}
	2 \pi^2 \, \di s^2 = -\frac{N(\tau)^2}{q(\tau)} \di \tau^2 + \frac{p(\tau)}{4} \left( \sigma_1^2 + \sigma_2^2 \right) + \frac{q(\tau)}{4} \sigma_3^2 \,,
\end{equation}
where $\sigma_1$, $\sigma_2$ and $\sigma_3$ are the left-invariant one-forms of SU(2) given by $\sigma_1 = -\sin\psi \dd\theta+\cos\psi \sin \theta \dd\phi, \sigma_2=\cos\psi \dd\theta+\sin\psi \sin \theta \dd\phi$ and $\sigma_3=\dd\psi+\cos \theta \dd\phi$, with $0 \leq \theta \leq \pi$, $0\leq \phi \leq 2\pi$ and $0 \leq \psi < 4\pi$. The state of the universe is specified by a wave function $\Psi (p, q)$ where the scale factors $p$ and $q$ are the two minisuperspace coordinates. In the parametrization \eqref{metricansatz} surfaces of constant $\tau$ are squashed three-spheres. The amount of squashing is conveniently expressed in terms of a parameter $\alpha \equiv p/q -1$; the round sphere corresponding to $\alpha=0$. The semiclassical quantum cosmology of this minisuperspace (with a particular focus on the no-boundary proposal) was previously studied in Ref. \cite{Jensen1991} (see also \cite{HAWKING198483,WRIGHT1985115,PhysRevD.31.3073}). In this paper we extend this work beyond the semiclassical approximation.

In the above parametrization of the metric the Einstein-Hilbert action can be written in phase space form as \cite{supplemental}
\begin{equation} \label{pqactionH}
	S[x,\Pi;N] = \int_0^1 \di \tau \left( \Pi_\alpha \dot{x}^\alpha - N H \right) \,,
\end{equation}
up to the appropriate boundary terms, where $x^\alpha \equiv (p,q)$, $\Pi_\alpha \equiv (\Pi_p, \Pi_q)$ are the momenta conjugate to $p$ and $q$, and
\begin{equation} \label{pqhamiltonian}
	H = \Pi_q \frac{q}{p} \Pi_q - 2 \Pi_q \Pi_p + \frac{q}{p} + \Lambda p - 4 \,.
\end{equation}
In \eqref{pqhamiltonian} a factor ordering is suggested which, upon canonical quantization of the system in position space, gives rise to a Laplacian ordering of the derivatives. This ensures that the quantization scheme is invariant under changes of the minisuperspace coordinates \cite{HalliwellWdW1988}. That is,
\begin{align}
	H &= - \frac{\hbar^2}{2} \nabla^2 + \frac{q}{p} + \Lambda p - 4 \label{HMSS} \,.
\end{align}
In the quantum theory wave functions $\Psi$ are annihilated by the operator version of the classical constraint $H = 0$ leading to the Wheeler-DeWitt equation, $H \Psi = 0$. Path integral solutions of the Wheeler-DeWitt equation may be obtained starting from a standard quantum-mechanical propagator between initial and final data in fixed time $N$ and then integrating $N$ over some contour, as described in more detail in the next section.

The Hamiltonian \eqref{pqhamiltonian} is linear in the coordinate $q$ and the momentum $\Pi_p$ which implies that the quantum system is exactly soluble. The propagator in position space is obtained by direct evaluation of the phase space path integral
\begin{equation} \label{pospropagator}
	K(x_1,N;x_0,0) = \hspace{-1mm} \underset{\hspace{-4mm}x(0) = x_0}{\overset{\hspace{3mm}x(1) = x_1}{\int}} \hspace{-3mm} \mathcal{D}x^\alpha \, \mathcal{D}\Pi_\alpha \, e^{iS/\hbar}
\end{equation}
since $q$ and $\Pi_p$ act as Lagrange multipliers, enforcing the classical equations of motion upon the remaining variables. (In \eqref{pospropagator} and the following, $N$ is a constant.) Thus the semiclassical ``approximation'' to \eqref{pospropagator} is exact, and it is straightforward to show that
\begin{equation} \label{pospropagator2}
	K(x_1,N;x_0,0) = \frac{1}{4 \pi \hbar \, N} \sqrt{\frac{p_0 p_1}{p_0 p_1 - N^2}} ~ e^{i S_1 / \hbar} \,,
\end{equation}
with
\begin{equation}
	S_1 = N \left[ 4 - \frac{\Lambda}{3} \left(\sqrt{p_0 p_1 - N^2} + p_0 + p_1 \right) \right] + \frac{1}{N} \left[ (q_0 + q_1) \sqrt{p_0 p_1 - N^2} - p_0 q_0 - p_1 q_1 \right] \,.
\end{equation}
An unambiguous definition of \eqref{pospropagator2} involves specifying its analytic structure in terms of $N \in \mathbb{C}$, which comes down to a choice of branch cut for the square root $\sqrt{p_0 p_1 - N^2}$. We return to this matter below. Note that the Schr\"odinger equation $H K = i \hbar \, \partial_N K$ is solved by \eqref{pospropagator2}, and that $\lim_{N \rightarrow 0} K(x_1,N;x_0,0) = \delta(x_1 - x_0)$ as appropriate for the position space propagator. We may calculate the propagator in any other representation by Fourier transformation, e.g.
\begin{equation} \label{mompropagator1}
	K(p_1,q_1,N;p_0,\Pi_{q,0},0) = \frac{1}{2 N} \sqrt{\frac{p_0 p_1}{p_0 p_1 - N^2}} ~ \delta \left( \Pi_{q,0} - \frac{p_0 - \sqrt{p_0 p_1 - N^2}}{N} \right) \, e^{i S_0 / \hbar} \,,
\end{equation}
where
\begin{equation} \label{S0}
S_0 = \frac{\Lambda \Pi_{q,0}}{3} N^2 + \left[ 4- \frac{\Lambda}{3} \left( 2 p_0 + p_1 \right)  \right] N - \Pi_{q,0} q_1 + \frac{q_1(p_0 - p_1)}{N} \,.
\end{equation}

Several other anisotropic minisuperspaces such as the Bianchi type I and III and Kantowski-Sachs models studied in e.g. Ref. \cite{Halliwell1990} also turn out to be exactly soluble. We will elaborate and exploit this feature elsewhere \cite{us2018}. It goes without saying that the exact solvability of these models is a feature of the minisuperspace truncation. Furthermore, a general truism about minisuperspace models is that the fluctuation determinant accompanying the exponential factor in a semiclassical approximation to a path integral is not robust with respect to the inclusion of other degrees of freedom. Going beyond minisuperspace could qualitatively alter the off-shell analysis of the path integral.

\section{No-Boundary wave function}\label{NBWFsec}
\noindent The original no-boundary proposal was not born fully-formed. Instead the intuitively appealing path integral construction has been developed and refined over many years. In the minisuperspace \eqref{pqactionH}-\eqref{pqhamiltonian} and in the gauge $\dot{N} = 0$ the no-boundary wave functions involve expressions of the following form \cite{HalliwellWdW1988,Halliwell:1990qr},
\begin{equation} \label{NBWF}
\Psi (y) = \sum_\mathcal{M} \int_\mathcal{C}\dd N \hspace{-3mm} \overset{\hspace{3mm}x(1) = y}{\int_\mathcal{B}} \hspace{-3mm} \mathcal{D}x^\alpha \, \mathcal{D}\Pi_\alpha ~ e^{i S[x,\Pi;N]/\hbar} ~.
\end{equation}
There is a family of wave functions implementing the no-boundary idea but differing in the choice of four-manifolds $\mathcal{M}$ in the sum in \eqref{NBWF}, in the boundary conditions $\mathcal{B}(\mathcal{M})$ at $\tau = 0$ on the lapse-dependent path integrals over $x$ and $\Pi$, and in the contours $\mathcal{C}(\mathcal{M})$ for the ordinary integrals over lapse values \cite{HarHal1990}. 

However the obvious requirement that the integral in \eqref{NBWF} converges and that the resulting wave function fits in a clear predictive framework for quantum cosmology, including the condition it be normalizable under an appropriate inner product\footnote{An example of such a normalization condition is the induced inner product, reviewed e.g. in Refs. \cite{Hartle:1997dc,PhysRevD.80.124032}.}, significantly limits the possible choices $\mathcal{M}$, $\mathcal{B}$ and $\mathcal{C}$ \cite{HarHal1990,Louko1988}. We now specify these features to define a NBWF which, we will show, obeys these basic criteria and whose predictions agree in the saddle point approximation with those of the original Hartle-Hawking NBWF.

First, the relevant four-manifolds in the no-boundary proposal are those with a single boundary and which admit everywhere regular `saddle point' solutions to the Einstein equation. In the minisuperspace \eqref{metricansatz} these are the closed four-ball $\overline{B^4}$, the complex projective plane with an open four-ball removed $\mathbb{C}\text{P}^2 \setminus B^4$ and the cross-cap $\mathbb{R}\text{P}^4 \setminus B^4$ \cite{PhysRevD.42.2458,Daughton:1998aa}. The regular solutions on the first two manifolds are (part of) the known Taub-NUT-de Sitter and Taub-Bolt-de Sitter solutions respectively. They are candidate saddle points of the above path integral. Here we concentrate on the contribution of the $\overline{B^4}$ topology only. Preliminary evidence indicates that including the other topologies does not significantly change our results \cite{us2018}.

\noindent All regular Taub-NUT-de Sitter solutions are of the form \eqref{metricansatz} with $p(\tau), q(\tau) \sim \pm 2iN_s \, \tau$ as $\tau \rightarrow 0$, where $N_s$ is one of a number of values for the lapse which enforces the Hamiltonian constraint $H = 0$ on solutions to the equations of motion. This behavior near the origin of the disk corresponds to the following conditions on the minisuperspace positions and momenta at $\tau = 0$: $p(0) = 0 = q(0), \Pi_p(0) = \mp i = \Pi_q(0)$. In the quantum theory only certain pairs of these classical conditions should be elevated to boundary conditions $\mathcal{B}$ on the path integral \eqref{NBWF} (excluding those pairs in which both a position and its momentum are fixed, which would be quantum-mechanically inconsistent). Here we adopt the following boundary conditions:
\begin{equation} \label{Bchoice}
\mathcal{B}: ~~~ p(0) = 0 \ , \quad \Pi_q(0) = -i \,.
\end{equation}
We will discuss alternative boundary conditions $\mathcal{B}$ in Ref. \cite{us2018} where we will argue that \eqref{Bchoice} are essentially the unique boundary conditions which yield a well-defined and normalizable NBWF in Bianchi IX minisuperspace. Note that the choice of sign for $ \Pi_q (0)$ in \eqref{Bchoice} will turn out to be crucial in obtaining a physically meaningful state.

Eq. \eqref{mompropagator1} shows that the propagator $K(p_1,q_1,N;p_0,\Pi_{q,0},0)$ contains an $N$-dependent delta function constraint. Therefore the boundary conditions \eqref{Bchoice} are singular at face value. To get around this problem, recall that we obtained the propagator in this mixed representation by Fourier transforming the position space propagator \eqref{pospropagator2}, which generates the delta function in \eqref{mompropagator1}. To implement the boundary conditions \eqref{Bchoice} we therefore perform a Laplace transform (cf. Refs. \cite{Halliwell1988,Halliwell1990}). That is, we convolve the position space propagator \eqref{pospropagator2} with $\exp(i \Pi_{q,0} q_0 / \hbar)$ and integrate $q_0$ over a half-infinite line. This yields the reciprocal of the argument of the delta function instead of the delta function itself. In the resulting object, one can show that the joint limit $(p_0 , \Pi_{q,0}) \rightarrow (0,-i)$ with constant ratio $\sqrt{p_0}/(\Pi_{q,0} + i)$ is well-defined.\footnote{The analytic structure of $\sqrt{p_0 p_1 - N^2}$ is important in this. We choose $\underset{p_0 \rightarrow 0}{\lim} \sqrt{p_0 p_1 - N^2} = + i N$. With the sign convention used in \eqref{S0}, and in the limit $(p_0 , \Pi_{q,0}) \rightarrow (0,-i)$, this choice renders the prefactor finite and the action equal to the one of an instanton with boundary data \eqref{Bchoice} at $\tau = 0$ and $(p(1),q(1)) = (p_1,q_1)$ at $\tau = 1$.} Taken together these specifications \textit{define} what we mean by the lapse-dependent path integral in \eqref{NBWF} in this model. We obtain
\begin{equation} \label{Ktilde}
	\overset{\hspace{3mm}x(1) = (p,q)}{\int_\mathcal{B}} \hspace{-3mm} \mathcal{D}x^\alpha \, \mathcal{D}\Pi_\alpha ~ e^{i S[x,\Pi;N]/\hbar} \propto \frac{\sqrt{p}}{N^2} \, e^{i S_0 / \hbar} \,,
\end{equation}
where $S_0$ is given by Eq. \eqref{S0}, with $p_1 = p, q_1 = q, p_0 = 0, \Pi_{q,0} = -i$.

\noindent Even though \eqref{Ktilde} is not obviously a propagator in the usual sense it is nevertheless an exact solution to the Schr\"odinger equation.\footnote{Indeed the Laplace transform is but one particular example of a class of linear transformations one can do on the initial data of a propagator that preserves its quality of solving the Schr\"odinger equation.} Moreover it takes on a semiclassical form with the action given by a regular instanton satisfying boundary data fitting to the no-boundary proposal. From a practical viewpoint the above manipulations simply serve to find an appropriate prefactor to accompany the semiclassical exponential factor specified by the no-boundary instanton. That is, a prefactor that guarantees the Wheeler-DeWitt equation is eventually satisfied.

Finally we turn to the contour $\mathcal{C}$ for the lapse in \eqref{NBWF}, for the topology $\mathcal{M} = \overline{B^4}$. We do not attribute much fundamental physical meaning to a particular choice of lapse contour in a given minisuperspace model, since examples show that the result obtained from any given choice can depend on the variables retained and even on the parametrization of the metric \cite{PhysRevD.43.2730}. Our contour choice is guided instead by the physically motivated and broadly applicable prescription given in Ref. \cite{HarHal1990} which, in this particular model, is conveniently implemented by a closed contour encircling the origin $N = 0$.\footnote{Closed contours in the context of the NBWF have been considered before (see e.g. Refs. \cite{Halliwell1988,Halliwell1989,Halliwell1990,HarHal1990}). Together with infinite contours they provide the only evident ways of generating wave functions constructed as functional integrals.} Other contours will be considered in Ref. \cite{us2018} and shown not to yield physically reasonable results in this model.

With this all elements pertaining to the $\overline{B^4}$ contribution to the NBWF are in place, and we get 
\begin{equation} \label{PsiHHpq}
	\Psi(p,q; \overline{B^4}) = \sqrt{p} \, \oint \di N \, \frac{1}{N^2} \exp\left\{ \frac{i}{\hbar} \left[ - \frac{i \Lambda}{3} N^2 + \left( 4 - \frac{\Lambda p}{3} \right) N + i q - \frac{p q}{N} \right] \right\} \,.
\end{equation}
A closed contour $\mathcal{C}$ ensures the wave function \eqref{PsiHHpq} satisfies the Wheeler-DeWitt equation exactly. Its semiclassical behavior is specified by the regular Taub-NUT saddle point solutions on $\overline{B^4}$ as we discuss further below.

\section{Damped perturbations}\label{HH}
\noindent Using the residue theorem one can express \eqref{PsiHHpq} as an infinite series. However it is illuminating to evaluate \eqref{PsiHHpq} in the semiclassical limit and in the large three-volume regime where the wave function describes an ensemble of classical histories. The saddle points $N_s$ of the exponent -- the lapse values which enforce the Hamiltonian constraint on the instantons -- are solutions of $2i\Lambda N_s^3 / 3 + \left( \Lambda p / 3 - 4 \right) N_s^2 - p q =0 \,.$ One of the three saddle points always lies on the positive imaginary axis. From \eqref{PsiHHpq} it follows that the semiclassical exponential factor associated with this saddle is purely real. If the wave function were dominated by this saddle point, it would not predict the universe to behave classically at large volume \cite{HHH2008}. Thus the contour should avoid a contribution from this saddle on physical grounds \cite{HarHal1990}. In the region of superspace
\begin{equation} \label{complexcondition}
\Lambda q > \frac{(\Lambda p)^2}{81}\left(\frac{12}{\Lambda p}-1\right)^3~,
\end{equation}
the two other saddles are complex and located symmetrically around the imaginary axis in the lower half part of the lapse plane \cite{supplemental}. The closed contour $\mathcal{C}$ we have chosen can be deformed into a sum of steepest descent contours which pick up the two complex saddle points only. The corresponding instantons belong to the Taub-NUT-de Sitter family and have a complex nut parameter \cite{us2018}.

In the large volume regime $p \gg 1/\Lambda$, with the ratio $p/q = 1 + \alpha$ finite, a straightforward calculation shows that
\begin{equation}\label{WF}
\Psi_\text{HH}(p, \alpha; \overline{B^4}) \propto \sqrt{\hbar} \Lambda \left( \frac{1+\alpha}{\Lambda p} \right)^{3/4} \exp\left[ \frac{6 (1+2\alpha)}{\hbar \Lambda (1+\alpha)^2} \right] \cos \left[ \frac{6}{\hbar \Lambda \sqrt{1+\alpha}} \left( \frac{\Lambda p}{3} \right)^{3/2} - \frac{3 \pi}{4} \right]
\end{equation}
to leading order in $1/ \Lambda p$. The asymptotic wave function \eqref{WF} satisfies the classicality conditions \cite{HHH2008,Hartle:2007gi}, $\left|\nabla \text{Re}(i \bar{S}_0)\right| / \left|\nabla \text{Im}(i \bar{S}_0)\right| \sim (\Lambda p)^{-3/2} \rightarrow 0 \text{ as } \Lambda p \rightarrow \infty \,,$ where $\bar{S}_0 \equiv S_0(N_s)$. This means it predicts an ensemble of classical histories that are anisotropic deformations of asymptotic de Sitter space. 
The classical asymptotic scale factors behave as $\Lambda p(t) = (1+\alpha) t$ and $\Lambda q(t) = t$. Therefore the individual histories can be labeled by the squashing $\alpha$ of their future (conformal) boundary. 

\begin{figure}[H]
\begin{center}
\includegraphics[scale=.45]{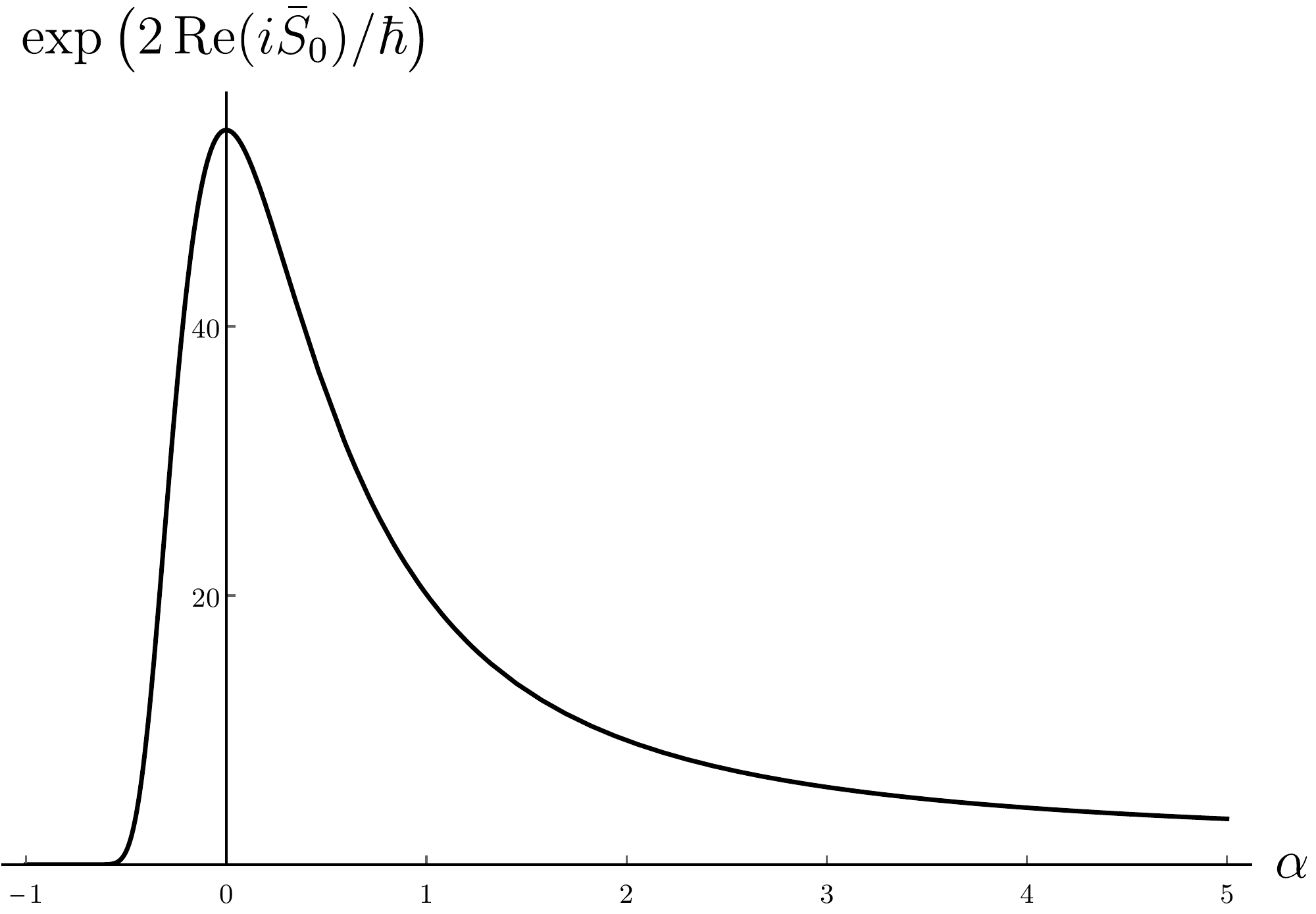}
\end{center}
\caption{\small The leading order in $\hbar$ probability distribution specified by \eqref{WF} over a one-parameter family of anisotropic deformations of de Sitter space labeled by the squashing parameter $\alpha$ of the future boundary. The NBWF predicts that small and large fluctuations away from isotropy ($\alpha = 0$) are suppressed. The values $\hbar = 1, \Lambda = 3$ were taken in \eqref{WF} for this plot.}\label{Onshell}
\end{figure}

The wave function \eqref{WF} specifies the leading order in $\hbar$ probabilities over histories. We show this in Figure \ref{Onshell} as a function of $\alpha$. The relative probabilities are typical of the Hartle-Hawking NBWF: the distribution is Gaussian around the isotropic de Sitter space.\footnote{The semiclassical exponent in Eq. (4.2) reduces to that of the NBWF in the minisuperspace considered in Refs. \cite{Halliwell1988,FLT1,DiazDorronsoro2017} on the isotropic $p=q$ slice. This agreement extends outside the large volume regime \cite{us2018} and includes the locations of the two complex saddles in the $N$-plane.} Large anisotropies with $q \gg p$ have $\alpha$ close to $-1$ and are exponentially suppressed. For large anisotropies $q \ll p$, i.e. large positive $\alpha$, we also see exponential suppression. For sufficiently large $\alpha$ the exponential suppression flattens out, but we expect on general grounds that the exact solution for the state is normalizable in the induced inner product for all $\alpha$ \cite{us2018}.

\section{Discussion}
\noindent We have shown there exists an implementation of the no-boundary idea expressed in terms of a gravitational path integral in an anisotropic minisuperspace model that yields a well-defined (normalizable) state in which deviations from isotropy are damped. The no-boundary proposal thus predicts that our universe should be isotropic with high probability.

The model we have studied -- the biaxial Bianchi IX minisuperspace -- is a non-linear completion of the minisuperspace spanned by a scale factor and the $n = 2$ gravitational wave mode perturbation of de Sitter space considered in Refs. \cite{FLT2,FLT3}. In those papers it is claimed that all no-boundary proposals are ill-defined due to problems with large perturbations. Our work disproves this claim. The discrepancy between our results and those of Refs. \cite{FLT2,FLT3} can be traced to two key features of the off-shell analysis. 

First, the analysis in Refs. \cite{FLT2,FLT3} is plagued by the breakdown of perturbation theory. This is because the integrand of the integral over the lapse in \eqref{NBWF} is non-analytic in perturbation theory. The authors of Refs. \cite{FLT2,FLT3} have included the off-shell contributions to the path integral associated with this non-analytic structure. This led them to conclude that fluctuations around isotropy are enhanced. However it turns out these contributions are an artefact of perturbation theory. Working with a non-linear completion of the theory we have shown that the integrand of the lapse integral is analytic everywhere, and hence that the above off-shell contributions are absent.

Second, the authors of Refs. \cite{FLT2,FLT3} implement the no-boundary idea by imposing the initial boundary condition that all variables go to zero both on-shell and off-shell. In a path integral representation of the no-boundary proposal this choice of boundary conditions gives rise to saddle point contributions in which the Euclidean lapse $N_E$ is negative for small geometries, thereby rendering the Euclidean action for fluctuations $\phi$ about those saddle points negative. This means the fluctuation wave functions are of the form $\exp ( + \phi^2)$ for small $\phi$, in stark contrast to the expected Bunch-Davies vacuum state wave functions and likely rendering the state ill-defined. This phenomenon was previously noticed in Refs. \cite{Halliwell1990,HarHal1990}.

We have instead implemented the no-boundary idea by requiring the three-volume to go to zero initially in combination with a specific regularity condition on the momentum of one of the variables. In particular we have imposed $\Pi_q(0) = - i$. If we had adopted the initial condition $\Pi_q(0) = +i$, the closed contour for the lapse would have selected the ``wrong sign'' saddle points discussed in Refs. \cite{FLT1,FLT2,FLT3}, leading to an unphysical exponentially growing behaviour. With our choice of sign the wrong sign saddle points are nowhere to be found, nor is any off-shell structure relevant to the semiclassical wave function. Instead we recover the original Hartle-Hawking NBWF which is normalizable and predicts that the amplitude of large anisotropies is strongly suppressed. A similar conclusion holds for gravitational wave and scalar field perturbations of de Sitter space with higher quantum numbers \cite{us2018}.

More generally our results suggest that a more fundamental implementation of no-boundary initial conditions in the isotropic minisuperspace model is not, as is traditionally done, to set the initial scale factor to zero, but instead to impose a semiclassically equivalent regularity condition on the momentum (as considered in \cite{Halliwell1990}). This also motivates more general investigations of the role of momentum boundary conditions in the NBWF, both initially and on the final boundary.\footnote{We thank E. Witten for correspondence on this.}

Finally we note that holography (or dS/CFT) postulates an alternative formulation of the wave function not in terms of a gravitational path integral but rather involving the partition function of dual (Euclidean) field theories defined directly on the final boundary \cite{Strominger:2001pn,Horowitz:2003he,Hertog2011,Anninos2012}. Our results qualitatively agree with recent holographic calculations of the NBWF in vector toy models defined on squashed three-spheres \cite{Anninos2012,Bobev:2016sap,Hawking:2017wrd}. This suggests that holography implements the specific no-boundary conditions that we made explicit here. It would be interesting to understand this aspect of the holographic dictionary in more detail.

\vskip .3cm
\noindent{\bf Acknowledgements:} We thank Alice Di Tucci, Job Feldbrugge, Ted Jacobson, Jean-Luc Lehners, Jorma Louko, Neil Turok and Alex Vilenkin for stimulating discussions. TH and YV are supported in part by the National Science Foundation of Belgium (FWO) grant G092617N, the C16/16/005 grant of the KULeuven and by the European Research Council grant no. ERC-2013-CoG 616732 HoloQosmos. JDD is supported by the National Science Foundation of Belgium (FWO) grant G.0.E52.14N Odysseus. OJ acknowledges support from the James Arthur Fellowship.

\bibliographystyle{klebphys2}
\bibliography{references}

\section{Supplementary Material}
\beginsupplement
\noindent In this addendum we provide supporting arguments for the main text. The commentary below is organized according to the sections of the main text and should be read in the context of the paragraph where we have redirected the reader to this document.

\setcounter{section}{1}
\section{}
\noindent In configuration space form the action reads
\begin{equation}
	S[x;N] = \int_0^1 \di \tau \, N \left( \frac{1}{2 N^2} f_{\alpha \beta}(x) \dot{x}^\alpha \dot{x}^\beta - U(x) \right) \,, \tag{S2.1}
\end{equation}
up to the appropriate boundary terms, where have defined the minisuperspace metric and potential
\begin{equation} \label{fMSS}
   f =
  \frac{-1}{2} \left( {\begin{array}{cc}
   q/p & 1\\
   1 & 0 \\
  \end{array}} \right) \,, \quad U = \frac{q}{p} + \Lambda p - 4 \,. \tag{S2.2}
\end{equation}
The momenta $\Pi_\alpha \equiv f_{\alpha \beta} \dot{x}^\beta / N$ are given explicitly by
\be \label{momenta}
\Pi_p = -\frac{1}{2N} \left( \frac{q \dot{p}}{p} + \dot{q} \right) \ , \quad \Pi_q = -\frac{1}{2N} \, \dot{p} \,. \tag{S2.3}
\ee 
We note that in general the Hamiltonian \eqref{HMSS} may contain another term, proportional to $\hbar^2$ times the scalar curvature on minisuperspace, with a proportionality coefficient such that $H$ is conformally invariant. This ensures that the quantization procedure is invariant with respect to redefinitions of the lapse function as well \cite{HalliwellWdW1988}. At the level of the phase space path integral such a term can be generated by a particular covariant skeletonization of the path integral \cite{Kuchar83}. In two-dimensional minisuperspaces this term is absent.

\setcounter{section}{3}
\section{}
\begin{figure}[H]
\begin{center}
\subfloat[]{\includegraphics[scale=0.40]{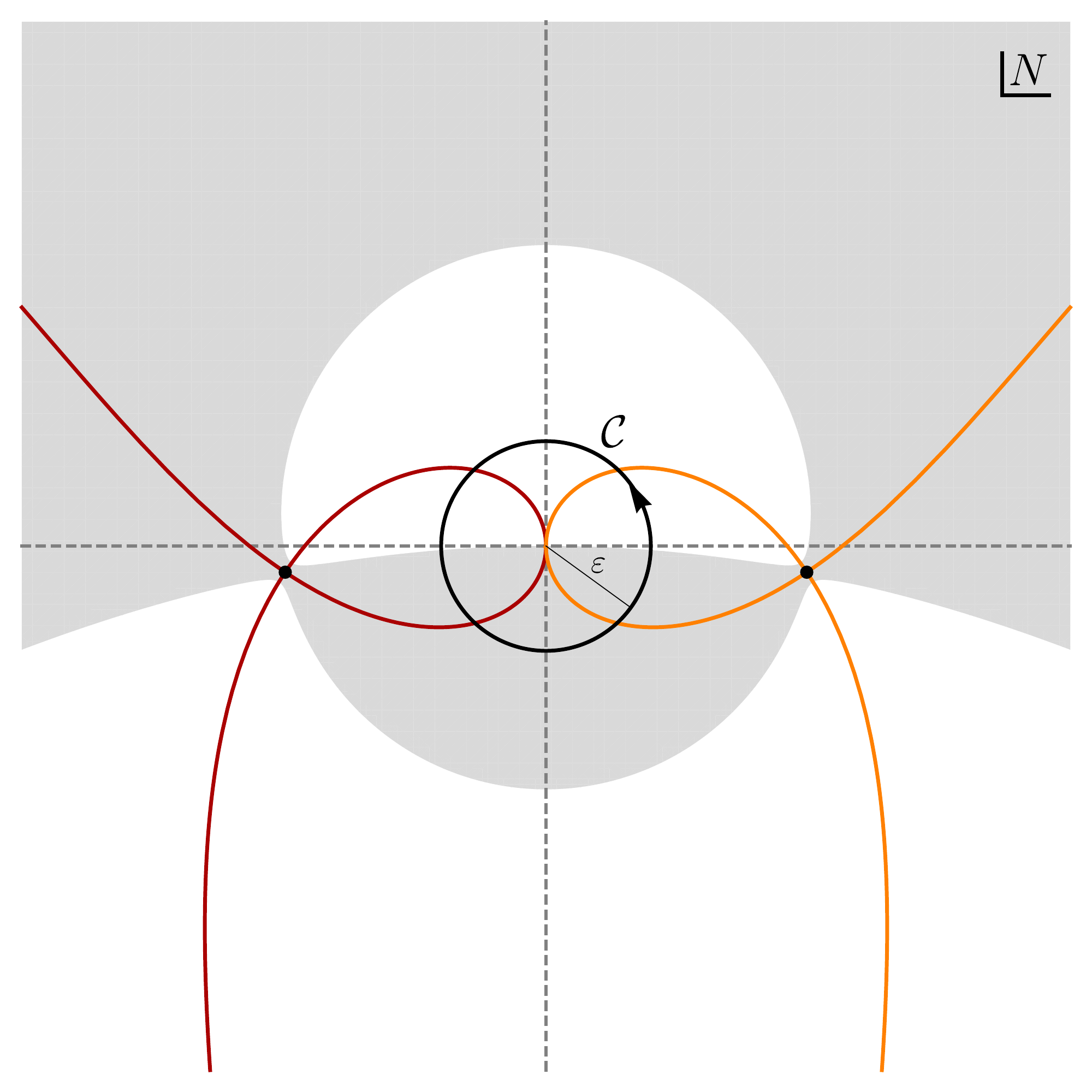}} \hfil
\subfloat[]{\includegraphics[scale=0.40]{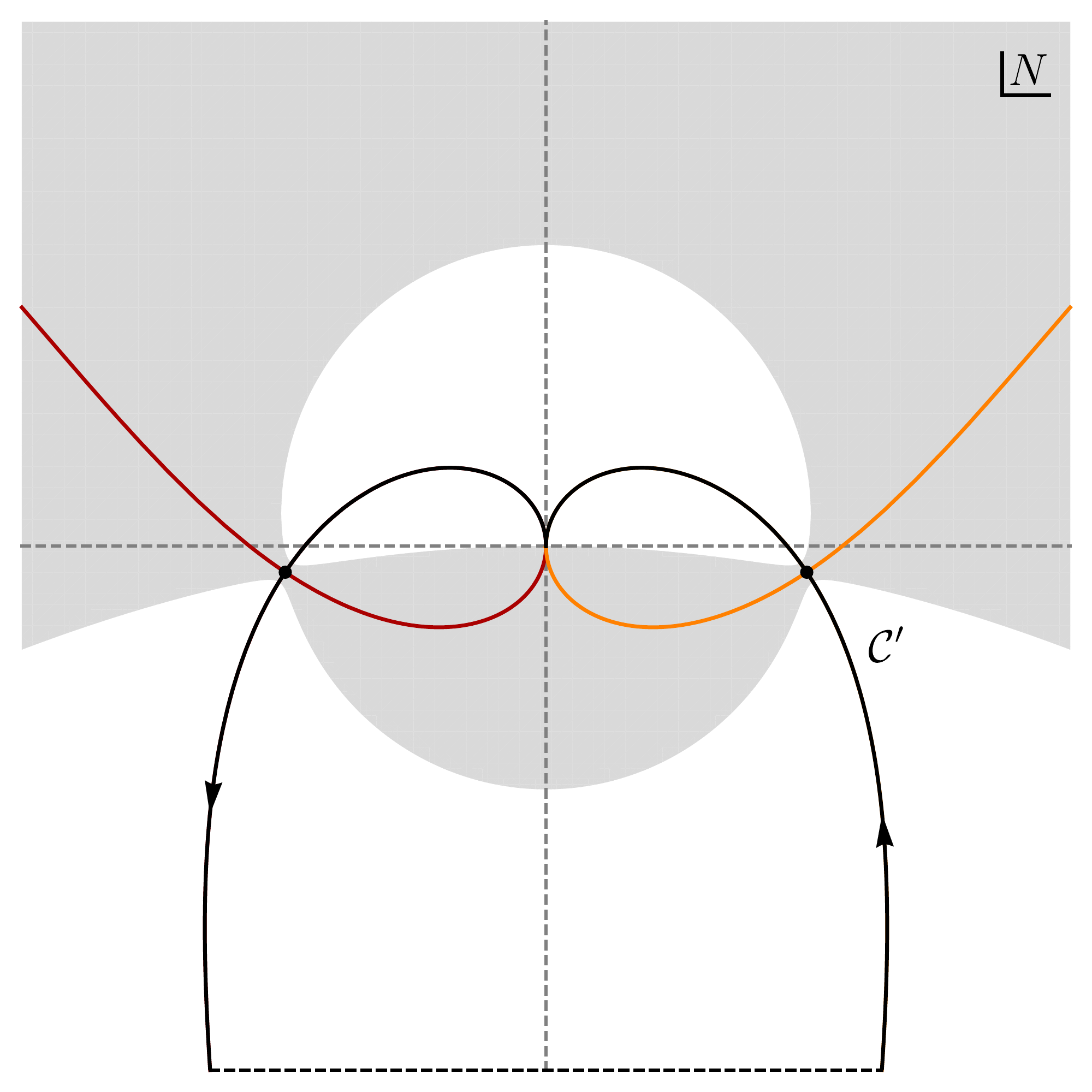}}
\end{center}
\caption{\small Two saddle points of $\text{Re}(i S_0)$ appearing in the semiclassical evaluation of the Hartle-Hawking NBWF, Eq. \eqref{PsiHHpq}, in biaxial Bianchi IX minisuperspace. The saddle points are shown as black dots in the complex $N$-plane together with their steepest ascent and descent curves. In the shaded regions $\text{Re}(i S_0) \geq 0$ while in white regions $\text{Re}(i S_0) < 0$. The third saddle lies on the positive imaginary axis and is not shown since it is irrelevant. The integral defining this NBWF involves a closed contour $\mathcal{C}$ around the origin (panel (a)). Its continuous deformation $\mathcal{C}'$ onto a sum of steepest descent contours (panel (b)) includes contributions from the two lower saddle points only. It follows from Eq. \eqref{PsiHHpq} that the descent lines asymptote to the negative imaginary axis, making the deformation valid. The numerical values $\Lambda = 3, p = 100$ and $\alpha = 0$ were taken to produce this figure. We find a qualitatively similar saddle point structure for all $\alpha$ in the large volume domain of this minisuperspace (defined by Eq. \eqref{complexcondition}).\label{PL-plots2}}
\end{figure}

\noindent The relevant saddle points are located at
\begin{equation}\label{Ns}
	\Lambda N_s = \pm \left[ \sqrt{\frac{3}{1+\alpha}} \, \sqrt{\Lambda p} + \mathcal{O}\left( \frac{1}{\sqrt{\Lambda p}} \right) \right] - i \left[ \frac{3}{1+\alpha} + \mathcal{O}\left( \frac{1}{\Lambda p} \right) \right] \,. \tag{S4.1}
\end{equation}
The action $S_0$ evaluated on these saddle points is
\begin{equation}\label{HH-saddle}
\Lambda \bar{S}_0 \equiv \Lambda S_0(N_s)  = \mp \left[ \sqrt{\frac{4}{3(1+\alpha)}} \, (\Lambda p)^{3/2} + \mathcal{O}\left( \sqrt{\Lambda p} \right) \right] - i \left[ \frac{6 (1+2\alpha)}{(1+\alpha)^2} + \mathcal{O}\left( \frac{1}{\Lambda p} \right) \right] \,, \tag{S4.2}
\end{equation}
and its second derivative is
\begin{equation}
	\frac{S_0''(N_s)}{\Lambda} = \mp \left[ 2 \sqrt{\frac{1+\alpha}{27}} \, \sqrt{\Lambda p} + \mathcal{O}\left( \frac{1}{\sqrt{\Lambda p}} \right) \right] - i \left[ \frac{8}{3} + \mathcal{O}\left( \frac{1}{\Lambda p} \right) \right] \,, \tag{S4.3}
\end{equation}
while the angle of the descent curves with the positive real $N$-axis at the saddles is given by
\begin{equation}\label{angle}
	\theta_s = \pm 3\pi/4 + \mathcal{O}\left( \frac{1}{\sqrt{\Lambda p}} \right) \,. \tag{S4.4}
\end{equation}
A steepest descent analysis of the remaining 1D integral that defines the Hartle-Hawking NBWF, Eq. \eqref{PsiHHpq}, is given in Figure \ref{PL-plots2}.

\end{document}